# Ferroelectricity in Polar ScAlN/GaN Epitaxial Semiconductor Heterostructures

Joseph Casamento[1, a)], Ved Gund[2], Hyunjea Lee[2], Kazuki Nomoto[2], Takuya Maeda[2], Benyamin Davaji[2], Mohammad Javad Asadi[2], John Wright[1], Yu-Tsun Shao[3] David A. Muller[3,4], Amit Lal[2], Huili (Grace) Xing[1,2,4], and Debdeep Jena[1,2,4]

[1]Department of Materials Science and Engineering, Cornell University, Ithaca, NY 14853, USA

[2]School of Electrical & Computer Engineering, Cornell University, Ithaca, NY 14853, USA

[3]School of Applied and Engineering Physics, Cornell University, Ithaca, New York 14853, USA

[4]Kavli Institute at Cornell for Nanoscale Science, Cornell University, Ithaca, New York 14853, USA

a) Corresponding author: jac694@cornell.edu

**Abstract**

Room temperature ferroelectricity is observed in lattice-matched ~18% ScAlN/GaN heterostructures grown by molecular beam epitaxy on single-crystal GaN substrates. The epitaxial films have smooth surface morphologies and high crystallinity. Pulsed current-voltage measurements confirm stable and repeatable polarization switching in such ferroelectric/semiconductor structures at several measurement conditions, and in multiple samples. The measured coercive field values are $E_c$~0.7 MV/cm at room temperature, with remnant polarization $P_r$~10 μC/cm$^2$ for ~100 nm thick ScAlN layers. These values are substantially lower than comparable ScAlN control layers deposited by sputtering. Importantly, the coercive field of MBE ScAlN is smaller than the critical breakdown field of GaN, offering the potential for low voltage ferroelectric switching. The low coercive field ferroelectricity of ScAlN on GaN heralds the possibility of new forms of electronic and photonic devices with epitaxially integrated ferroelectric/semiconductor heterostructures that take advantage of the GaN electronic and photonic semiconductor platform, where the underlying semiconductors themselves exhibit spontaneous and piezoelectric polarization.

Ferroelectrics are a relatively rare class of materials that possess the ability to reversibly switch their electrical polarization (P) with an applied electric field (E). This phenomenon, coupled with the already existing piezoelectric and pyroelectric behavior present in crystals with lower crystal symmetry, garners tremendous attention in research and technological applications. [1-4] Specifically, attention has been focused on the downsizing of electronic systems such as piezoelectric actuators, radiofrequency (RF) filters for higher frequency operation, and the development of subthreshold transistors for low voltage applications. [5] The hysteresis in the polarization and electric field relationship in a ferroelectric material allows for memory functionality. Namely, the two stable polarization states in a ferroelectric (e.g., up and down) can be stored as on and off states in an electronic device, respectively. When a ferroelectric layer can be integrated with a semiconductor, a merging of the memory and logic functions are possible. [6]

The existing III-nitride semiconductor materials - GaN, AlN and InN (and their corresponding alloys) boast desirable optoelectronic features such as direct band gaps tunable from the infrared to ultraviolet regime, pyroelectric behavior, and novel heterostructures that form the basis of polarization engineering and polarization-induced doping. [7] Accordingly, the nitrides have found rapidly expanding applications in solid-state lighting, RF and power electronics, and piezoelectric actuators and detectors. [8-13] For example, GaN based LEDs and Lasers are used extensively in visible and short-wavelength photonics, and GaN based high-electron mobility transistors (HEMTs) form the basis for high frequency RF amplifiers and power electronics. Likewise, AlN is the current material of choice for piezoelectric applications due to high figures of merit stemming from coupled piezoelectric and mechanical behavior. [14-16]

Until very recently, ferroelectric behavior had not been seen in the III-nitride material family. [17] The energy to switch between the stable metal and nitrogen-polar states in the wurtzite crystal was expected to be larger than the energy for dielectric breakdown. Alloying the III-nitrides with transition metals such as scandium (Sc) and yttrium (Y) with predicted large solubility in the wurtzite crystal structure is expected to increase the piezoelectric response and induce ferroelectric response by increasing the ionic bond character and decreasing the energy barrier for polarization, respectively. This has been proven experimentally in sputtered $Sc_xAl_{1-x}N$ films, with a large piezoelectric enhancement of over 400% for Sc contents up to 40%, and robust ferroelectric behavior for Sc contents as low as 10% and as thin as 20 nm. [18-22] These developments in conjunction with the relatively low temperatures required for sputter deposition, have led to $Sc_xAl_{1-x}N$ usage as a complementary metal oxide semiconductor (CMOS) compatible element and its current widespread commercial production. [23,24]

To exploit the semiconducting properties of GaN and the ferroelectric properties of the $Sc_xAl_{1-x}N$ materials system, their epitaxial integration is the first step. To this end, molecular beam epitaxy (MBE) aims to accomplish this with epitaxial growth on semiconductor bulk single-crystals with low dislocation densities. In addition, the usage of an ultra-high vacuum environment allows for low impurity levels in the resulting $Sc_xAl_{1-x}N$ films. It is scientifically intriguing to contrast the ferroelectric properties of sputter-deposited films with MBE grown ones that can be integrated with high electron mobility transistors (HEMTs) and light emitting diodes (LEDs). For such devices, low defect densities, low impurity levels, and ultra-thin $Sc_xAl_{1-x}N$ layers are necessary for optimal performance. GaN, finding widespread usage in these optoelectronic applications, as well as being in-plane lattice-matched to $Sc_xAl_{1-x}N$ at ~18% Sc, is an ideal platform to epitaxially stabilize and study the fundamental properties of wurtzite $Sc_xAl_{1-x}N$. [25]

In this work, we report the epitaxial growth of $Sc_xAl_{1-x}N$-GaN heterostructures and the observation of robust room temperature ferroelectric behavior at a nominal Sc composition of 18% Sc. In-situ reflection high-energy electron diffraction (RHEED) images for $Sc_xAl_{1-x}N$ indicate epitaxial growth for all layers and suggest that $Sc_xAl_{1-x}N$ maintains a wurtzite crystal structure when grown on top of GaN. X-Ray Diffraction (XRD) shows a strong $Sc_xAl_{1-x}N$ 002 peak, indicative of a wurtzite crystal structure and high crystalline quality. Atomic force microscopy (AFM) images show smooth surface morphologies from epitaxial growth with sub nanometer rms roughness. Pulsed IV measurements at a variety of temperatures and frequencies confirm polarization switching and ferroelectric behavior occurs before electrical conduction (e.g., leakage) starts to dominate at higher applied voltages. The relatively low spontaneous polarization and coercive field values compared to sputter deposited $Sc_xAl_{1-x}N$ control samples indicate several potential technological applications of such epitaxial ferroelectric/semiconductor heterostructures.

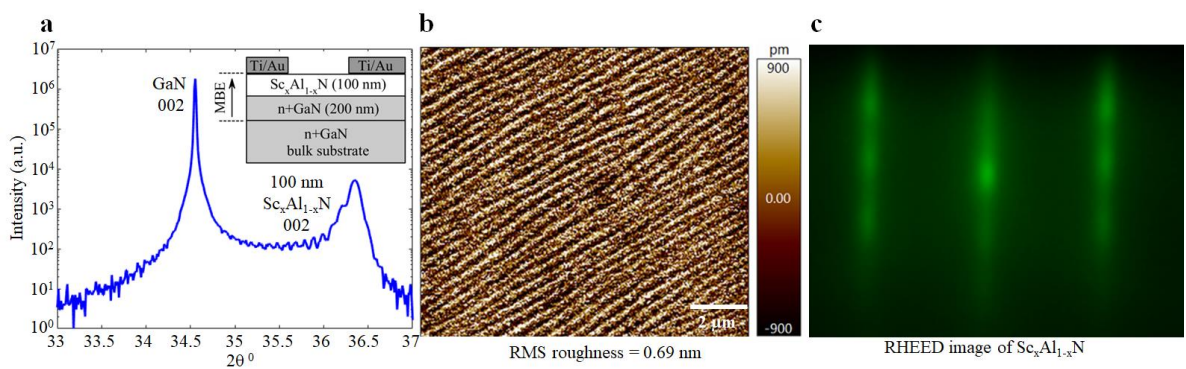

**Figure 1**: (a) Symmetric XRD 2theta-omega scans, showing a strong wurtzite 002 $Sc_xAl_{1-x}N$ peak for 100 nm thick $Sc_xAl_{1-x}N$. The sample layer structure is shown in the inset. (b) AFM image and (c) RHEED image along the [110] zone axis of $Sc_xAl_{1-x}N$, showing relatively smooth surface morphologies and epitaxial growth on GaN.

The Sc$_x$Al$_{1-x}$N/GaN layer structures studied this work were grown by plasma-assisted MBE and processed into structures shown in the inset of **Figure 1 (a)**. The MBE-grown doped n+ GaN layer under the Sc$_x$Al$_{1-x}$N layer forms the bottom electrode, and the top electrodes were deposited after the growth. Details of the epitaxial process and measurement and characterization techniques are described in the methods section. The XRD scan of the heterostructure in **Figure 1 (a)** shows a strong wurtzite Sc$_x$Al$_{1-x}$N peak with thickness fringes indicative of a smooth interface. The AFM image shown in **Figure 1 (b)** indicates a smooth surface morphology with a rms roughness of 0.69 nm on a 10x10 μm scale that result from the epitaxial growth on a single-crystal substrate. The in-situ RHEED image shown in **Figure 1 (c)** indicates epitaxial growth on GaN of single-crystalline Sc$_x$Al$_{1-x}$N, with primary streak azimuths preserved along the [110] zone axis. Combined, these data suggest wurtzite Sc$_x$Al$_{1-x}$N with no signs of secondary phases and crystal misorientations. Spots observed along the primary streaks in RHEED indicate a relatively three-dimensional growth mode, which is expected for nitrogen-rich growth conditions needed for the growth of smooth, and highly crystalline Sc$_x$Al$_{1-x}$N. [26,27]

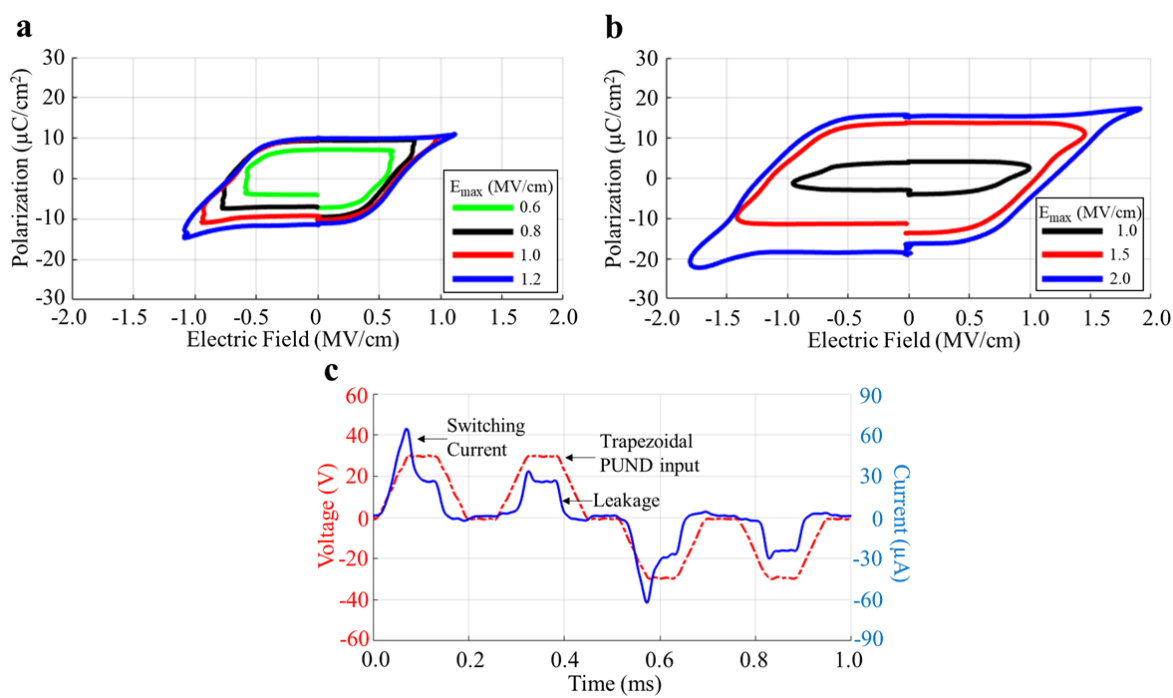

**Figure 2**: (a,b) P-E loops for 100 nm and 200 nm thick Sc$_x$Al$_{1-x}$N / n$^+$ GaN heterostructures, showing hysteresis loops and convergence of remnant polarization and coercive field values at higher applied fields. (c) A representative PUND waveform using trapezoidal voltage pulses on the 200 nm sample, indicating the ability to separate the contributions of polarization switching and electrical leakage to the measured current. Similar behavior is seen for triangular voltage pulses.

**Figures 2 (a) and (b)** show the measured polarization-electric field (P-E) loops for the $Sc_xAl_{1-x}N$/GaN structures where the thickness of the $Sc_xAl_{1-x}N$ layers are 100 nm (Fig 2a) and 200 nm (Fig 2b). As shown in **Figure 2 (c)** for the 200 nm $Sc_xAl_{1-x}N$/GaN structure, the polarization switching current is observed in the P and N pulses and is calculated by subtracting the currents in the U or D pulses, respectively. The U and D pulses are referred to as the "non-switching" pulses and they do not contain any switching current for a stable ferroelectric material. For a ferroelectric that is completely switched (e.g., very few or no misoriented domains exist), the calculated remnant polarization should remain constant as the electric field value is increased. The trend seen in **Figure 2** is observed for all the $Sc_xAl_{1-x}N$ samples that show ferroelectricity. At low electric fields, the hysteresis loops do not "close", indicative of incomplete switching. The calculated remnant polarization values and measured coercive field also do not converge to a consistent value for the same reason. At higher electric fields, the remnant polarization and coercive field converge to consistent values of ~10 $\mu C/cm^2$ and 0.7 MV/cm and ~15.5 $\mu C/cm^2$ and 1.2 MV/cm for the 100 nm and 200 nm thick $Sc_xAl_{1-x}N$ films, respectively. This is indicative of ferroelectric switching. Overall, these values and trends are consistent among several electrodes in each sample as discussed later, confirming repeatable ferroelectric behavior.

The measured remnant polarization and coercive field values are significantly lower than those measured in sputter-deposited $Sc_xAl_{1-x}N$ films. [17-22,28,29] A comparative plot is shown later. This may arise for several reasons. One reason is that the sputter deposited films have significantly different lattice constants and corresponding c/a ratios than the MBE grown $Sc_xAl_{1-x}N$ films. This leads to differences in the cation-anion c-axis bond length and the corresponding energy landscape of ferroelectric switching. In addition, the microscale and nanoscale defect morphologies of MBE grown $Sc_xAl_{1-x}N$ and sputter deposited $Sc_xAl_{1-x}N$ can be different, which would impact the switching dynamics and coercive field. Another point of note is that the maximum obtainable electric fields are limited by electrical leakage currents and are not as high as the sputter deposited films. Therefore, the measured polarization may be underestimated and may change if a higher electric field range is accessible.

The ferroelectric coercive field is a function of frequency, similar to the dielectric permittivity, via the Ishibashi Orihara model. [30] A simple analogy is that at high frequency, the atoms cannot respond to the applied voltage (electric field), and the measured coercive field is higher than at a lower frequency. The fundamental ferroelectric switching speed in proper ferroelectrics is related to the acoustic phonon velocity in the crystal and is typically much faster than the frequency ranges measured.

In addition, frequency dependent measurements are a means to assess potential trapping/de-trapping behavior of trap states inside the bandgap of a semiconducting material. The trapping and de-trapping process is also a transient phenomenon, but unlike polarization switching, the amount of trapped charge is a function of frequency, whereas the amount of polarization switching charge is not. The measured switching current density will increase as frequency increases as the charge has to be switched in a shorter amount of time, but the polarization charge is constant. Charge trapping and de-trapping is also time-dependent, and if present, would manifest as subsequent voltage pulses resulting in different current levels. That is not observed in these samples with identical current response in subsequent "UU" and DD" voltage pulses, as shown in **Figure 3**, confirming ferroelectric switching behavior.

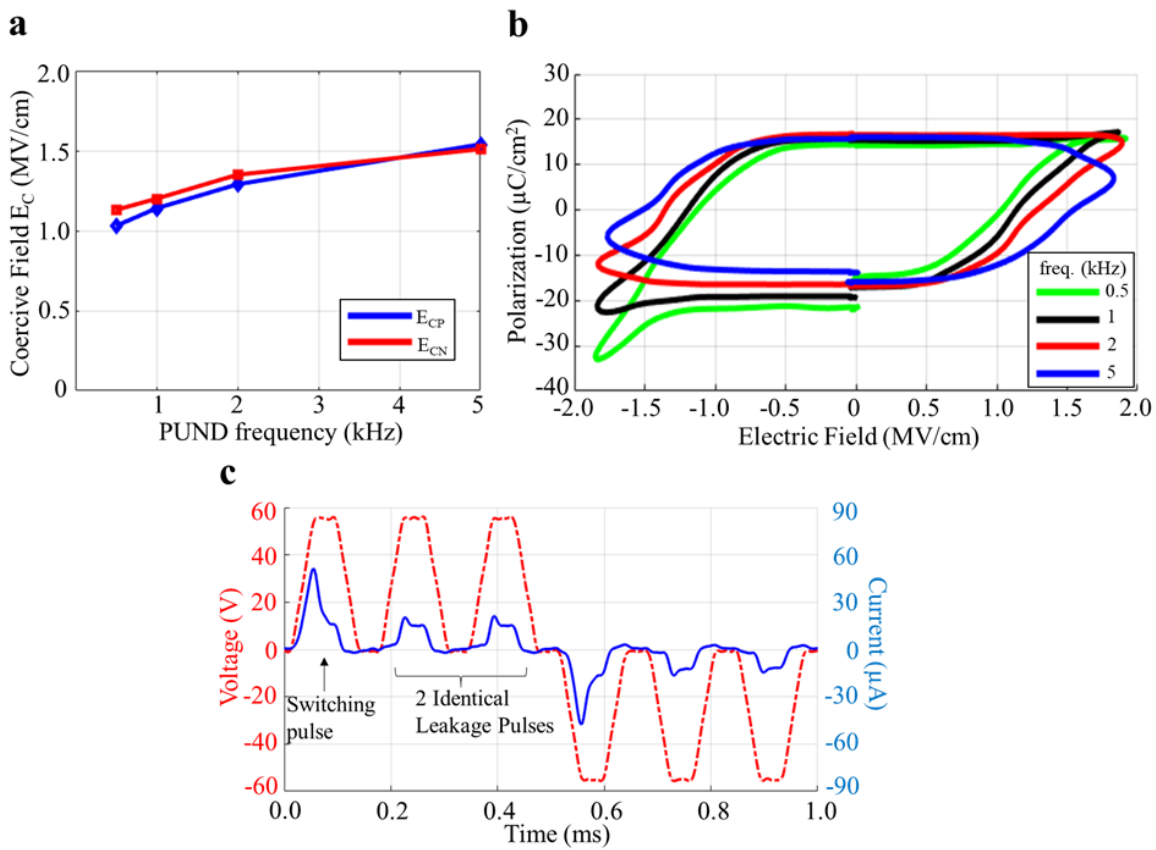

**Figure 3**: 200 nm thick $Sc_xAl_{1-x}N$ / $n^+$ GaN heterostructure with (a) measured coercive field as a function of measurement frequency, showing an increase in coercive field as measurement frequency increases. (b) P-E loops at various measurement frequencies, showing an increasing coercive field and a constant remnant polarization. (c) I-V waveform in the PUUNDD sequence, where subsequent pulses indicate identical behavior, verifying the peak from the first pulse is from ferroelectric switching.

In **Figure 3** it is noted that the P-E loop shapes broaden at higher frequencies, indicative of the RC time constant of the system. Namely, the system cannot dissipate the charge in the measurement time at higher frequencies. The same behavior is observed for the 100 nm thick $Sc_xAl_{1-x}N$ / $n^+$ GaN sample.

The ferroelectric coercive field is also a function of temperature. The coercive field is an empirical value and is extrinsic in that it incorporates nucleation and growth of domains at defects and preferred orientation sites in conventional ferroelectric capacitors, though the case could differ for a single-crystal epitaxial structure. The measured coercive field is generally much less than that predicted by Landau-Ginzburg theory. [31,32] Generally, the coercive field should decrease as temperature increases due to increased thermal energy for domain wall motion and nucleation which reduces the barrier for switching. [33] PUND measurements at various temperatures on the MBE $Sc_xAl_{1-x}N$ / $n^+$ GaN heterostructures confirm this, as shown in **Figures 4 (a) and (b)** for the 200 nm $Sc_xAl_{1-x}N$ / $n^+$ GaN sample. The coercive field monotonically decreases as the temperature is increased. The same behavior is observed for the 100 nm $Sc_xAl_{1-x}N$ / $n^+$ GaN sample. Details about the specific temperature measurement setup and calibration are described elsewhere. [34] The data is repeatable over many measurement cycles and also over several devices, as illustrated in **Figure 5.**

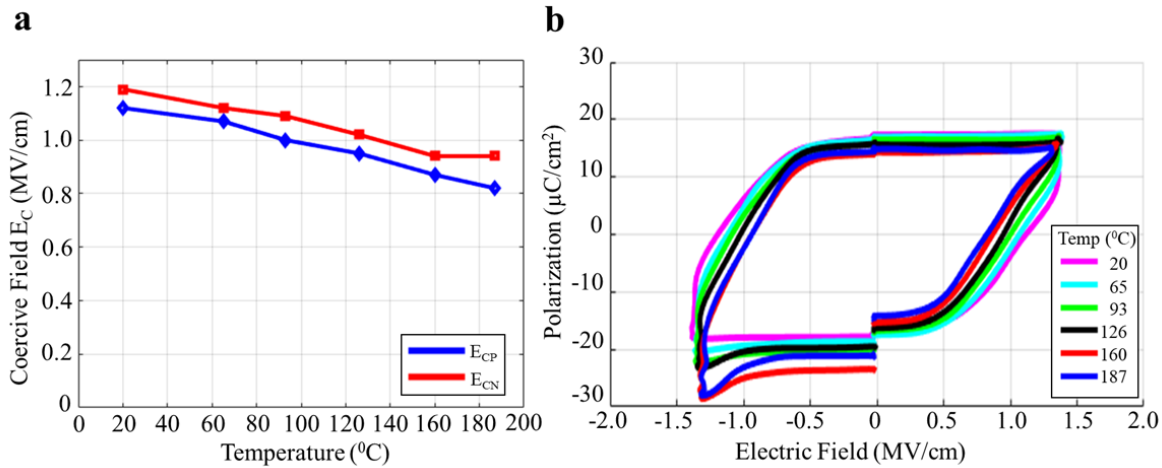

**Figure 4**: Measured coercive field (a) for 200 nm thick $Sc_xAl_{1-x}N$ / $n^+$ GaN sample and extracted P-E loops (b) as a function of measurement temperature. The coercive field monotonically decreases with temperature and the remnant polarization is constant.

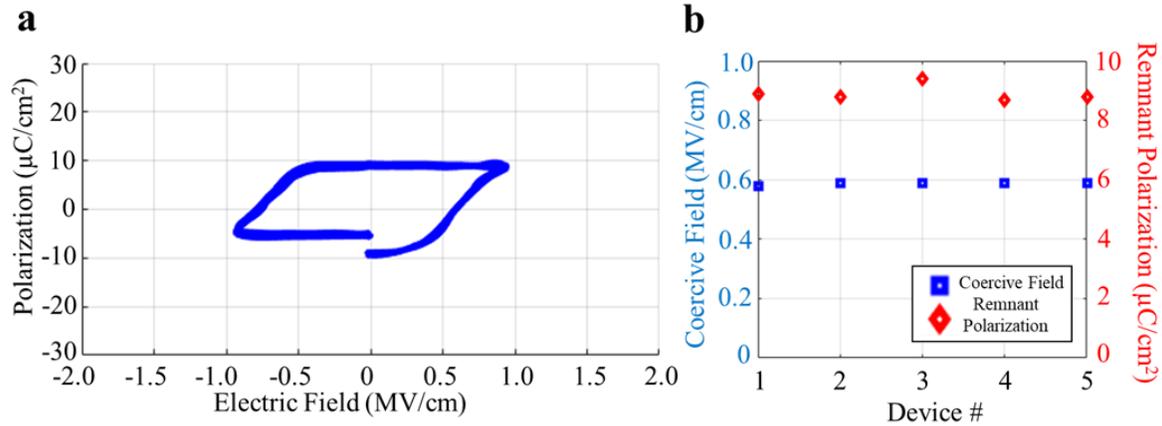

**Figure 5**: (a) Overlaid P-E loops of 200 nm thick $Sc_xAl_{1-x}N$ / $n^+$ GaN sample, showing repeatability over many measurement cycles (b) Measured $E_c$ and $P_r$ values over different electrode combinations on the sample, indicating consistent results

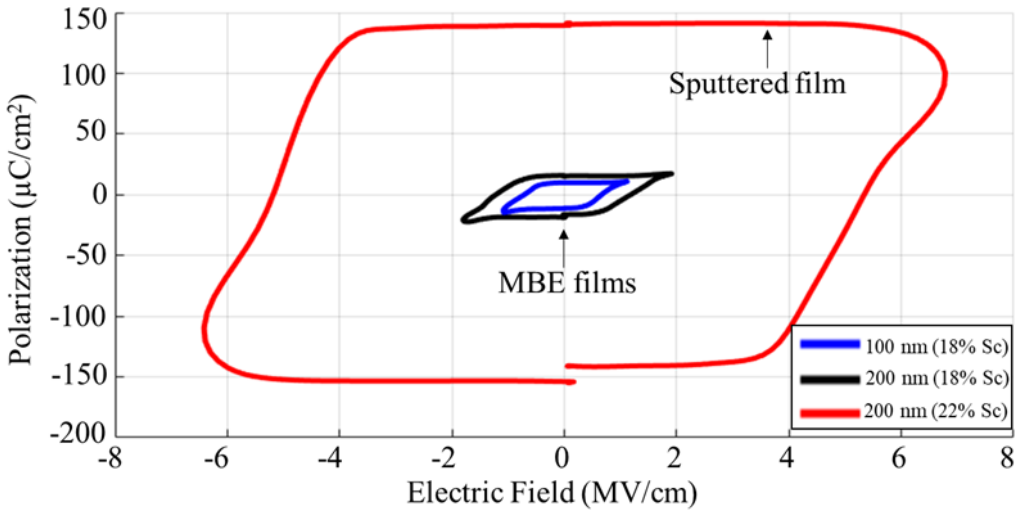

**Figure 6**: Overlay of the room temperature P-E loops of sputtered 200 nm $Al_{0.78}Sc_{0.22}N$ with MBE 200 nm and 100 nm $Al_{0.82}Sc_{0.18}N$. The MBE films have a much lower $P_r*E_c$ product, indicating their suitability in low voltage operation.

As pointed out earlier, the $Sc_xAl_{1-x}N$ films grown by MBE show significantly reduced coercive fields and remnant polarization values compared to the sputter-deposited films, as indicated in **Figure 6**. Table 1 summarizes the values as a comparative study between various $Sc_xAl_{1-x}N$ films measured with the same ST setup at Cornell.

**Table 1**: Comparison of sputter deposited and MBE-grown $Sc_xAl_{1-x}N$ measured at Cornell, with deposition method/thickness, electrode materials, Sc atomic percentage, coercive field ($E_c$) and remnant polarization ($P_r$) values given

| Deposition method and thickness | Bottom Electrode/Top Electrode | Sc% | $E_c$ (MV/cm) | $P_r$ (μC/cm$^2$) |
|---|---|---|---|---|
| Sputter (200nm) | 200 nm Pt/200 nm Mo | 22 | 5.3 | 139 |
| MBE (200 nm) | GaN/200 nm Au | 18 | 1.2 | 15.2 |
| MBE (100 nm) | GaN/200 nm Au | 18 | 0.65 | 10.0 |

In conclusion, ferroelectric behavior is observed in epitaxial MBE grown ScAlN samples of 100 and 200 nm thickness on n type GaN substrates. The measured coercive field and remnant polarization values of 10 μC/cm$^2$ and 0.7 MV/cm for the 100 nm thick $Sc_xAl_{1-x}N$ film at 1 kHz frequency and 300K are substantially lower than the typical range of values reported for sputter deposited $Sc_xAl_{1-x}N$ measured at similar frequencies and temperatures. Since the critical fields in sputtered $Sc_xAl_{1-x}N$ layers typically exceed the breakdown field of GaN (~3.5 MV/cm), the low coercive fields of epitaxial layers allow for device structures where the ferroelectricity and semiconducting properties can safely coexist. The measured coercive field decreases as temperature is increased and increases as frequency is increased, as expected. The P-E loops show saturation at moderate electric fields and are stable over many cycles, confirming ferroelectric response and not trapping behavior. This data points toward the utilization of $Sc_xAl_{1-x}N$ as an epitaxial ferroelectric in GaN based optoelectronic devices with a variety of potential applications ranging from reconfigurable logic, RF, and photonic devices, and most importantly bringing a natural epitaxially integrated memory element into the rapidly expanding wide-bandgap nitride semiconductor ecosystem. The fact that the underlying semiconductor layer GaN itself possesses strong spontaneous polarization, and also strong piezoelectricity in its combination with strained AlGaN, InAlN, and AlN layers, opens up a rich playground for the interplay of polarization in semiconductor physics in ways that have not been possible before.

**Acknowledgements**

This work was supported by the DARPA Tunable Ferroelectric Nitrides (TUFEN) program monitored by Dr. Ronald G. Polcawich. The work was also supported in part by NSF DMREF grant 1534303, Cornell's nanoscale facility (grant NCCI-1542081), AFOSR grant FA9550-20-1-0148, NSF DMR-1710298, and the Cornell Center for Materials Research Shared Facilities which are supported through the NSF MRSEC program (DMR-1719875). The authors would like to acknowledge Matt Besser and Trevor Riedemann and the Materials Preparation Center, Ames Laboratory, US DOE Basic Energy Sciences, Ames, IA, USA for supplying the Sc source material. The authors thank Dr. Susan Trollier-McKinstry of Pennsylvania State University, Dr. Darrell Schlom of Cornell University, and Drs. Ronald Polcawich, Brendan Hanrahan and Glen Fox of DARPA for illuminating discussions.

The authors declare that the data supporting the findings of this study are available within the paper and its supplementary information files.

**Methods**

The $Sc_xAl_{1-x}N$/GaN heterostructures of this work (**Figure 1a**) were grown by MBE in a Veeco® GenXplor system with a base pressure of ~$10^{-10}$ Torr on conductive n-type bulk GaN substrates for electrical measurements. Purified elemental Sc (from Ames Laboratory) of nominally 99.9% purity (including C and O impurities), Aluminum (99.9999% purity), gallium (99.99999% purity), and silicon (99.9999% purity) were supplied using Knudsen effusion cells. Nitrogen (99.99995%) active species were supplied using a Veeco® RF UNI-Bulb plasma source, with total chamber pressure of approximately $10^{-5}$ Torr during growth. The RF plasma power was kept at 200W and the total gas flow rate was 1.95 sccm. The growth temperature mentioned is the substrate heater temperature measured by a thermocouple. The homoepitaxial GaN layers were grown under standard metal-rich conditions at a temperature of ~595°C. This layer was doped heavily with the shallow donor dopant Silicon at a density of ~$2 \times 10^{19}$/cm$^3$ to form the bottom electrode. Sc and Al atomic percentages in the film were adjusted by the ratio of the respective fluxes from the effusion cells. For the $Sc_xAl_{1-x}N$ layers, Sc and Al were co-deposited under nitrogen-rich conditions with III/V ratio ~ 0.85 at a substrate temperature of ~495°C. Nitrogen-rich growth conditions were utilized to prevent any excess metal formation and direct reaction of Sc and Al. A more in-depth study of the justification of growth conditions and calibration to establish the effective III/V ratio is described elsewhere.[35]

In-situ monitoring of the surface crystal structure was performed using a KSA Instruments RHEED apparatus with a Staib electron gun operating at 15 kV and 1.5 A. X-Ray Diffraction (XRD) was performed on a PanAlytical Empyrean diffractometer at 45 kV, 40 mA with Cu Kα1 radiation (1.54057 Å). Post growth AFM measurements were performed using an Asylum Research Cypher ES system.

Pulsed electrical current-voltage (I-V) measurements in a positive-up, negative-down (PUND) waveform were performed. A Rigol DG1022 Arbitrary Waveform Generator was used to program the input signal, whose voltage was amplified with a PiezoSystems EPA-102 amplifier with 400V peak-to-peak output and 250 kHz bandwidth. A series resistor with an 11.6 Ω resistance was used to measure the switching currents. The resistor value was chosen to minimize the impact of the film as a voltage divider with the drive capacitor. Metal electrodes of Ti/Au were patterned lithographically to serve as the top contacts.

To test for ferroelectricity, a modified Sawyer-Tower (ST) setup was used.[36] More details about the setup are described elsewhere.[37] The films were tested with continuous wave positive-up-negative-down (PUND) input waveforms and both triangular and pulses of equal rise, fall, and wait times. The I-V data from the PUND waveform was utilized to construct polarization- electric field (P-E) loops in a lateral geometry (e.g., both electrodes on top of the sample) with circular electrodes of 40 and 400 μm respectively. The effective capacitance, formed by the two electrodes in series with the bottom n$^+$ GaN as the intermediate node is approximately equal to the capacitance formed between the 40 μm electrode and n$^+$GaN due to the large difference in the sizes of the metal electrodes. The polarization is given as the polarization switching current, which is the polarization charge divided by the area of the metal electrode. The polarization charge is calculated from the polarization switching current integrated over time. The switching current in the positive and negative voltage cycles, respectively, is identified as the total measured current minus the displacement current and any leakage currents. Displacement currents follow the form I = C dV/dt and leakage currents scale with increased voltage. The electric field is given as the applied voltage divided by the Sc$_x$Al$_{1-x}$N layer thickness.

# References

[1] J.Y. Kim, M.-J. Choi, and H.W. Jang, APL Mater. **9**, 021102 (2021).

[2] S. Das, Z. Hong, M. McCarter, P. Shafer, Y.-T. Shao, D.A. Muller, L.W. Martin, and R. Ramesh, APL Mater. **8**, 120902 (2020).

[3] T. Mikolajick, U. Schroeder, and S. Slesazeck, IEEE Trans. Electron Devices **67**, 1434 (2020).

[4] S. Trolier-McKinstry, Am. Ceram. Soc. Bull. **99**, 22 (2020).

[5] V. Fridkin and S. Ducharme, Ferroelectricity at the Nanoscale (Springer, 2014).

[6] J.F. Scott, Ferroelectric Memories (Springer, 2000).

[7] C. Wood and D. Jena, Polarization Effects in Semiconductors-From Ab Initio Theory to Device Applications (Springer, 2008).

[8] M.-A. Dubois, P. Muralt, and V. Plessky, in *1999 IEEE Ultrason. Symp. Proc. Int. Symp. Cat No99CH37027* (IEEE, Caesars Tahoe, NV, USA, 1999), pp. 907–910.

[9] M.D. Hodge, R. Vetury, S.R. Gibb, M. Winters, P. Patel, M.A. McLain, Y. Shen, D.H. Kim, J. Jech, K. Fallon, R. Houlden, D.M. Aichele, and J.B. Shealy, in *2017 IEEE Int. Electron Devices Meet. IEDM* (IEEE, San Francisco, CA, USA, 2017), p. 25.6.1-25.6.4.

[10] A. Hickman, R. Chaudhuri, S.J. Bader, K. Nomoto, K. Lee, H.G. Xing, and D. Jena, IEEE Electron Device Lett. **40**, 1293 (2019).

[11] S.M. Islam, K. Lee, J. Verma, V. Protasenko, S. Rouvimov, S. Bharadwaj, H. (Grace) Xing, and D. Jena, Appl. Phys. Lett. **110**, 041108 (2017).

[12] Z. Zhang, M. Kushimoto, T. Sakai, N. Sugiyama, L.J. Schowalter, C. Sasaoka, and H. Amano, Appl. Phys. Express **12**, 124003 (2019).

[13] V.J. Gokhale and M. Rais-Zadeh, J. Microelectromechanical Syst. **23**, 803 (2014).

[14] R. Ruby, P. Bradley, J. D. Larson, and Y. Oshmyansky, Electron. Lett. **35**, 794 (1999).

[15] R. Ruby, in *2011 Symp. Piezoelectricity Acoust. Waves Device Appl. SPAWDA* (IEEE, Shenzhen, China, 2011), pp. 365–369.

[16] R.H. Olsson, J.G. Fleming, K.E. Wojciechowski, M.S. Baker, and M.R. Tuck, in *2007 IEEE Int. Freq. Control Symp. Jt. 21st Eur. Freq. Time Forum* (IEEE, Geneva, Switzerland, 2007), pp. 412–419

[17] S. Fichtner, N. Wolff, F. Lofink, L. Kienle, and B. Wagner, J. Appl. Phys. **125**, 114103 (2019).

[18] M. Pirro, B. Herrera, M. Assylbekova, G. Giribaldi, L. Colombo, and M. Rinaldi, in *2021 IEEE 34th Int. Conf. Micro Electro Mech. Syst. MEMS* (IEEE, Gainesville, FL, USA, 2021), pp. 646–649

[19] D. Wang et al., *IEEE Electron Device Lett.* **41**, 1774 (2020).

[20] S. Yasuoka et al., *J. Appl. Phys.* **128**, 114103, (2020).

[21] M. Akiyama, T. Kamohara, K. Kano, A. Teshigahara, Y. Takeuchi, and N. Kawahara, Adv. Mater. **21**, 593 (2009).

[22] M. Akiyama, K. Kano, and A. Teshigahara, Appl. Phys. Lett. **95**, 162107 (2009).

[23] T. Mikolajick, S. Slesazeck, H. Mulaosmanovic, M.H. Park, S. Fichtner, P.D. Lomenzo, M. Hoffmann, and U. Schroeder, J. Appl. Phys. **129**, 100901 (2021).

[24] V.V. Felmetsger, in *2017 IEEE Int. Ultrason. Symp. IUS* (IEEE, Washington, DC, 2017), pp. 1–5.

[25] S. Leone, J. Ligl, C. Manz, L. Kirste, T. Fuchs, H. Menner, M. Prescher, J. Wiegert, A. Žukauskaitė, R. Quay, and O. Ambacher, Phys. Status Solidi RRL – Rapid Res. Lett. **14**, 1900535 (2020).

[26] M.T. Hardy, E.N. Jin, N. Nepal, D.S. Katzer, B.P. Downey, V.J. Gokhale, D.F. Storm, and D.J. Meyer, Appl. Phys. Express **13**, 065509 (2020).

[27] M.T. Hardy, B.P. Downey, N. Nepal, D.F. Storm, D.S. Katzer, and D.J. Meyer, Appl. Phys. Lett. **110**, 162104 (2017).

[28] K. Yazawa, D. Drury, A. Zakutayev, and G.L. Brennecka, Appl. Phys. Lett. **118**, 162903 (2021).

[29] X. Liu, J. Zheng, D. Wang, P. Musavigharavi, E.A. Stach, R.Olson III, and D. Jariwala. arXiv:2021.10019.

[30] H. Orihara, S. Hashimoto, and Y. Ishibashi, J. Phys. Soc. Jpn. **63**, 1031 (1994).

[31] V. L. Ginzburg, Zh. Eksp. Teor. Fiz. **15**, 739 (1945).

[32] S. Ducharme, V.M. Fridkin, A.V. Bune, S.P. Palto, L.M. Blinov, N.N. Petukhova, and S.G. Yudin, Phys. Rev. Lett. **84**, 175 (2000).

[33] W.J. Merz, Phys. Rev. **95**, 690 (1954).

[34] V. Gund, B. Davaji, H. Lee, M.J. Asadi, J. Casamento, H.G. Xing, D. Jena, and A. Lal, to appear in *IEEE ISAF 2021*.

[35] J. Casamento, C.S. Chang, Y.-T. Shao, J. Wright, D.A. Muller, H. (Grace) Xing, and D. Jena, Appl. Phys. Lett. **117**, 112101 (2020).

[36] C.B. Sawyer and C.H. Tower, Phys. Rev. **35**, 269 (1930).

[37] V. Gund, B. Davaji, H. Lee, J. Casamento, H.G. Xing, D. Jena, and A. Lal, to appear in *IEEE Transducers 2021*.